# Ultrafast Spin-to-Orbit and Orbit-to-Local-Spin Conversions of Tightly Focused Hybridly Polarized Light Pulses

*Yanxiang Zhang, Zijing Zhang\*, Zhongquan Nie\**

Dr. Y. Zhang, Prof. Z. Zhang

Department of Physics

Harbin Institute of Technology

Harbin 150001, China

**E-mail:** zhangzijing@hit.edu.cn

Dr. Z. Nie

Key Lab of Advanced Transducers and Intelligent Control System

Ministry of Education and Shanxi Province, College of Physics and Optoelectronics

Taiyuan University of Technology

Taiyuan, 030024, China

**E-mail:** niezhongquan1018@163.com

Spin-orbit interaction (SOI) have provided a new viable roadmap for the development of spin-based photonics devices. However, existing strategies to control the SOI focus commonly on tailoring the spatial dimension of light fields yet neglecting the inherent temporal one. Herein, we first present ultrafast temporal effects on both spin-to-orbit and orbit-to-local-spin conversions based on the time-assistant vectorial diffractive theory and the fast Fourier transformation. Such interconversions depend upon whether the incident hybridly vectorial light pulse carries vortex phase or not in a single high numerical aperture geometry. For the case of the absence of vortex phase, we find that it enables orbit angular momentum-carrying transverse component fields, and the resultant orbit angular momentum embedded within focused light fields remains constantly revolving as time elapses, which indicates that the controllable spin-to-orbit conversion occurs. By contrast, it is revealed that hybridly polarized vectorial-vortex light pulses allow access to the locally excited circular polarization at the focus, and this induced local circular polarization is independent of ultrafast varying time, whereas the resulting spin angular momentum density components experience the alternation between the appearance and annihilation over time, thus giving rise to the tunable orbit-to-local-spin conversion. These exotic ultrafast interconversions not only breathe a new life into the area of ultrafast photonics, but refresh our understanding on the paradigm of photonic SOI.

## 1. Introduction

Spin and orbit angular momenta (SAM and OAM), respectively, associated with the spatial polarization singularity and phase singularity of light wave, have appreciably enriched the avenue of light-matter interaction and propelled a myriad of fields and advanced techniques forward, such as topological photonics, chiral optics, nano-particle manipulation and spin-related beam shaping [1-4]. Here, the spatial polarization singularity is C-point at which the polarization state is left- or right-hand circular polarization but around which it is elliptical counterpart [5,6]. While the spatial phase singularity dictates null intensity pattern on the axis featured by annular light beam, namely, optical vortex, with the azimuthally associated spiral phase front mathematically manifested by the factor of $exp(il\varphi)$, where $l$ is a boundless integer referred to as topological charge (TC) or vortex order, and $\varphi$ denotes azimuthal coordinate [7]. In quantum scenarios, SAM and OAM possess the magnitude of $\sigma\hbar$ and $l\hbar$ per photon, respectively, in which $\sigma = \pm 1$ depends on circular polarization signed by $\pm$ for right- and left-handedness and $\hbar$ represents the reduced Plank constant. Their directions can be charactered by a moving particle spinning around its own axis for the former whereas orbiting the optical axis for the latter.

In particular, recent progresses on the internal conversion between SAM and OAM (i.e., spatial polarization and phase degrees of freedom of the light wave) have attracted widespread attentions owing to the fundamental importance and possible advantage to achieve the spin-Hall effects in inhomogeneous media, spin-directional coupling within evanescent near fields, and higher quantum dimensionality in quantum information [8,9]. On one hand, lots of approaches touching on the spin-to-orbit (STO) conversions, including tight focusing [10,11], nano-particle scattering [12,13], planar-surface reflection and refraction [14] as well as metamaterial designing [15,16], have been reported, making it convenient and feasible to bring novel functionalities into optical nano-devices. On the other hand, orbit-to-local-spin (OTLS) strategies, which can be successfully conversed from either the scalar-vortex fields or the vector-vortex fields into the circular polarization with left- or right- handedness, have been theoretically proposed [17-21] and experimentally substantiated [22,23]. Despite these impressive advances, there have never yet been involved the hybridly polarized light beams (including linear, circular and elliptical polarizations). This misses out plenty of opportunities offered by hybridly polarized light. What is more, such aforementioned twofold conversions can be usually achieved in merely a single-oriented fashion, i.e., either from spin to orbit conversion or the reverse, whereas the reciprocal conversions of the SAM and OAM, especially within the case of one kind of illumination light beams, remain far from satisfactory, due mainly to the scant attention paid to effective target light sources.

In addition, ultrafast light field control has been viewed as a crucial toolkit and played an important role on the implementation of ultrafast light-matter interaction. To date, a plethora of attempts have been proposed to tailor light wave beams or packets for the purpose of the on-demand generation and manipulation of specific structured light fields owing to urgent realistic requirements, such as subtle amplitude or phase encoding and retrieving, scalable spatial polarization modulation as well as sophisticated temporal shaping [24-29]. Among them, although the vigorous temporal wavefront shaping technologies broaden the degree of freedom towards controlling light fields, its multiplex ability is relatively weak. In this regard, active metamaterials consisting of manmade meta-atom building blocks can be regarded as a powerful tool to overcome this restriction owing to the fact that they can harness indiscriminate dimensions of the controllable space-time light fields by engineering the specific feature parameters of the nanostructure units [30]. Most recently, Zhan et al. demonstrated the existence of transverse OAM, namely space-time optical vortex, measured by the interference between a compressed femtosecond wave packet and picosecond ones reconfigured by the modified pulse shaper [31]. After that, this group also proposed another approach via tight focusing regime in order to generate transverse OAM [32]. As such, we have also predicted and interpreted in detail ultrafast time-varying phenomena within tightly focused light fields, especially for the ultrafast radially polarized light fields [33]. These impressive insights derive basically from the fundamental conception that space and time are tightly intertwining in Maxwell's equations. In spite of the advanced progresses on ultrafast light field control, there has been a key obstacle of the ultrafast spin-orbit interconversion, especially within the versatile hybridly polarized light pulses. Such significant and technological gaps give a challenge to incubating various roadmaps of the ultrafast light-matter interaction.

Motivated by these problems, we theoretically propose and demonstrate a fleetingly interconvertible strategy between SAM and OAM, based on the time-associated vectorial diffraction theory and the fast Fourier transform. We first examine that the incident hybridly vectorial light pulses (including linear, circular and elliptical polarizations) will unambiguously determine two OAM-carrying transverse component fields by exploiting a single high numerical aperture (NA) objective lens, which means that the STO conversion arises. This outcome is substantially distinct with conventional circularly polarized light fields. We further show that such focused light fields with resultant OAM remain constantly rotating as time elapses. Oppositely, it is revealed that the local circular polarization at the central focus can be induced by tightly focusing hybridly vectorial-vortex light pulses (including only linear polarization), thus implying that OTLS occurs. Such abductive local circular polarizations are independent of the ultrafast varying time, while there exists ultrafast temporal effect on three-orthogonal SAM density components. As an alternative way, by delicately tuning the values of distinct TC and tailoring corresponding time intervals, we are capable to readily harness the SAM density fields as desired functionalities. These proposed methodology and findings reported here might shed novel light on ultrafast science and facilitate the development of ultrafast light-matter interaction employing the additional temporal degree of freedom.

## 2. Theoretical Analyses of Ultrafast Hybridly Polarized Vectorial-Vortex Light Fields

**Figure 1** conceptually shows the schematic illustrations of our methodology with respect to ultrafast reciprocal conversions between SAM and OAM. As showcased, an incident hybridly polarized light pulse in the absence of vortex phase is firstly focused by an aplanatic high NA objective lens, thus resulting in the two OAM-carrying transverse field components (*x*- and *y*-field components) within tightly focused light field, in which ultrafast STO conversion occurs. One the other hand, the same hybridly polarized counterpart in the presence of vortex phases is also focused by the same objective lens, enabling the locally induced circular polarization at the focus of the total light field, therefore allowing access to the ultrafast OTLS conversion. In this connection, we start from an incident hybridly polarized light field with the vortex-dressed Laguerre-Gaussian (LG) femtosecond pulse envelope, which in the pupil plane can be featured by [34-36],

$$E(r,\varphi,t) = \left(\frac{\sqrt{2}r}{\sigma_0}\right)^{|l|} \exp\left(-\frac{r^2}{\sigma_0^2}\right) \exp(il\varphi) A(t) \mathbf{P}_h, \tag{1}$$

where ($r$, $\varphi$, $t$) represents the space-time coordinates in the cylindrical coordinate, in which $r = f\sin\theta$ meets the Abbe's sinusoidal condition [37] with the focal length $f$ of the objective lens, and $\varphi = arc\tan(y/x)$ in the initial plane. $\sigma_0$ stands for the light beam waist size. $A(t) = \exp[-(a_g t/T)^2]\exp(-i\omega_0 t)$ represents the temporal pulse with Gaussian envelope, wherein $a_g = (2ln2)^{1/2}$, $T$ is the pulse width of femtosecond laser light and $\omega_0$ is the center angular frequency of the carrier. Mathematically, the generalized vectorial polarization (GVP) can be charactered by the following formalism [38, 39],

$$\mathbf{P} = \frac{1}{\sqrt{2}}\left[\exp(-i\delta)\mathbf{e}_1 + \exp(i\delta)\mathbf{e}_2\right], \tag{2}$$

here, $\mathbf{e}_1$ and $\mathbf{e}_2$, respectively, denote a pair of basic vectors with the orthogonal polarization states. $\delta$ describes the spatially varying phase typically possessing arbitrary values. From Equation (2), when we select a pair of orthogonal linear polarization ($\mathbf{e}_x$ and $\mathbf{e}_y$) as the basic vector, the GVP can be further reduced to a so-called hybrid polarization state with locally linear polarization [40], described as,

$$\mathbf{P}_h = \frac{1}{\sqrt{2}}\left[\exp(i\delta)\mathbf{e}_x + \exp(-i\delta)\mathbf{e}_y\right], \qquad (3)$$

in which we cherry-pick $\delta = m\varphi + 2\pi nr/\sigma_0 + \varphi_0$ as the spatially varying phase, where $m$ and $n$ are azimuthal and radial index of the polarization, $\varphi$ and $r$ represent the azimuthal angle and polar axis under the cylindrical coordinates, and $\varphi_0$ is the initial polarized angle. After tedious derivation, we can revisit and deduce Equation (3) as the following formulism,

$$\mathbf{P}_h = \begin{bmatrix} P_x \\ P_y \\ P_z \end{bmatrix} = \frac{1}{2}\begin{bmatrix} \exp(i\delta)\left[\cos 2\varphi(\cos\theta-1)+\cos\theta+1\right]+\exp(-i\delta)\sin 2\varphi(\cos\theta-1) \\ \exp(i\delta)\sin 2\varphi(\cos\theta-1)-\exp(-i\delta)\left[\cos 2\varphi(\cos\theta-1)+\cos\theta+1\right] \\ 2\sin\theta\left(\exp(i\delta)\cos\varphi+\exp(-i\delta)\sin\varphi\right) \end{bmatrix}, \qquad (4)$$

From the analytical expression in Equation (4), it is indicated that the factor of $\exp(im\varphi)$ can always be regarded as common divisor for $P_x$ and $P_y$, thus leading to the expected orbit angular momentum within $x$- and $y$- component fields.

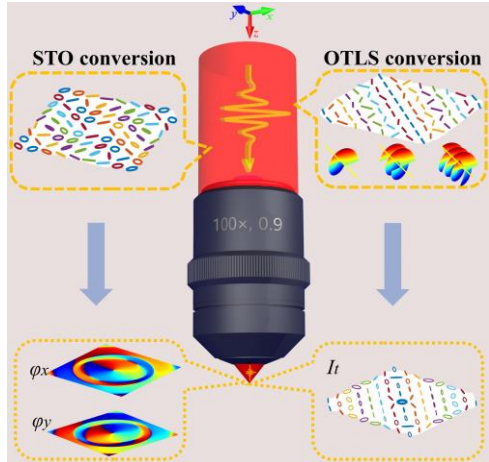

**Figure 1.** Schematic of ultrafast spin-orbit reciprocal conversion within tightly focused hybridly polarized light fields. The left side of the objective lens illustrates ultrafast STO conversion, while the right of that depicts ultrafast OTLS conversion. The former is performed by that the incident hybridly polarized LG femtosecond light pulse with $m = 1$, $n = 0$, $\varphi_0 = 0$, $l = 0$ (polarization map in the upper left inset) is focused through high NA objective lens, resulting in the two OAM-carrying transverse component fields (phase maps in the lower left inset) within focused light field. $\varphi_x$ and $\varphi_y$ are respectively the phase distributions of $x$- and $y$- component fields. The latter is implemented by which the incident hybridly vectorial-vortex analogous with $m = 0$, $n = 0$, $\varphi_0 = 0$, $l = 1, 2, 3$ (polarization map together with spiral phase fronts in the upper right inset) is focused through high NA objective lens, leading to the local circular polarization at the focus of the total fields $I_t$ (polarization map in lower right inset).

We then introduce the incident light fields charactered by Equation (1) into a high NA objective lens, after which a three-dimension (3D) polarized tightly focused light fields can be well-established. In this way, according to the time-dependent vectorial diffraction theory [34,35], such 3D light fields in optical spectrum domain near the focus point can be formulized as,

$$\mathbf{E}(r,\psi,z,\omega) = \frac{-if}{2\pi c}\int_0^\alpha \int_0^{2\pi} \omega S(r,\omega) F(r,\psi,z) \mathbf{P}(\theta,\varphi) \mathrm{d}\varphi \mathrm{d}\theta, \quad (5)$$

where $(r, \psi, z, \omega)$ is the cylindrical coordinate near the focus. Here, $\psi$ can be defined as the azimuthal angular with respect to $x$- axis, $\omega = kc = 2\pi nc/\lambda$ denotes the angular frequency of light pulse with wave vector of $k$, light velocity $c = 3 \times 10^8$ m/s in vacuum, the refractive index of the medium $n = 1$ for the free space used throughout this paper and $\lambda$ represents the center wavelength of the light pulses. $\alpha$ stands for the maximum convergence angle of the semi-aperture angular of $\theta$, determined by $\alpha = arc\sin(\mathrm{NA}/n)$. Besides, $F(r, \psi, z) = \sin\theta\cos^{1/2}\theta\exp[i\omega r\sin\theta\cos(\varphi-\psi)/c+i\omega z\cos\theta/c+il\varphi]$. $S(r, \omega)$ represents the apodization function of single optical spectrum component, manifested by the following form [34],

$$S(r,\omega) = \frac{1}{\sqrt{2\pi}}\int_{-\infty}^{\infty} E(r,t)\exp(i\omega t)dt = \frac{T}{2\sqrt{ln(2)}}\left(\frac{\sqrt{2}r}{\sigma_0}\right)^{|l|}\exp\left(-\frac{r^2}{\sigma_0^2}\right)\exp\left[-\frac{T^2(\omega-\omega_0)^2}{8ln(2)}\right]; \quad (6)$$

here, $E(r, t)$ refers to Equation (1). Submitting $S(r, \omega)$ together with Equation (4) into Equation (5), one can obtain the analytic expressions of the 3D orthogonally polarized field components in the optical spectrum domain after tedious calculation. Subsequently, on the basis of the inverse fast Fourier transform of each spectral components near the focal point,

$$E_j(r,\psi,z,t) = \mathcal{F}^{-1}\{E_j(r,\psi,z,\omega)\} \quad (j=x,y,z), \quad (7)$$

we can further deduce Equation (5) into the following focused light fields of hybridly polarized vectorial-vortex light pulses within the space-time domain, including $x$-, $y$- and $z$- polarized component fields,

$$E_x(r,\psi,z,t)$$
$$= -\frac{if}{2c}\int_0^\infty \int_0^\alpha F(r,z,\omega)\left\{\begin{array}{l}(\cos\theta+1)i^{m+l}J_{m+l}(\beta)\exp(i\Delta_1)+ \\ \frac{1}{2}(\cos\theta-1)\left[\begin{array}{l}i^{m+l+2}J_{m+l+2}(\beta)\exp(i(\Delta_1+2\psi))+ \\ i^{m+l-2}J_{m+l-2}(\beta)\exp(i(\Delta_1-2\psi))- \\ i^{-(m-l-3)}J_{-(m-l-2)}(\beta)\exp(-i(\Delta_2-2\psi))+ \\ i^{-(m-l+1)}J_{-(m-l+2)}(\beta)\exp(-i(\Delta_2+2\psi))\end{array}\right]\end{array}\right\}d\theta d\omega$$

$$E_y(r,\psi,z,t)$$

$$= -\frac{if}{2c}\int_0^\infty \int_0^\alpha F(r,z,\omega) \left\{ \begin{array}{l} (1+\cos\theta)i^{-(m-l)}J_{-(m-l)}(\beta)\exp(-i\Delta_2) + \\ \frac{1}{2}(1-\cos\theta)\begin{bmatrix} i^{-(m-l-2)}J_{-(m-l-2)}(\beta)\exp(-i(\Delta_2-2\psi)) + \\ i^{-(m-l+2)}J_{-(m-l+2)}(\beta)\exp(-i(\Delta_2+2\psi)) + \\ i^{m+l+3}J_{m+l+2}(\beta)\exp(i(\Delta_1+2\psi)) - \\ i^{m+l-1}J_{m+l-2}(\beta)\exp(i(\Delta_1-2\psi)) \end{bmatrix} \end{array} \right\} d\theta d\omega$$

$$E_z(r,\psi,z,t) = -\frac{if}{2c}\int_0^\infty \int_0^\alpha \sin\theta F(r,z,\omega) \begin{bmatrix} i^{m+l+1}J_{m+l+1}(\beta)\exp(i(\Delta_1+\psi)) + \\ i^{m+l-1}J_{m+l-1}(\beta)\exp(i(\Delta_1-\psi)) - \\ i^{-(m-l-2)}J_{-(m-l-1)}(\beta)\exp(-i(\Delta_2-\psi)) + \\ i^{-(m-l)}J_{-(m-l+1)}(\beta)\exp(-i(\Delta_2+\psi)) \end{bmatrix} d\theta d\omega , \quad (8)$$

wherein $\Delta_1 = (m+l)\psi + 2\pi n(f\sin\theta/\sigma_0)^2 + \varphi_0$ and $\Delta_2 = (m-l)\psi + 2\pi n(f\sin\theta/\sigma_0)^2 + \varphi_0$. $J_m(\beta)$ is the $m$-order Bessel function of the first kind with $\beta = kr\sin\theta$. $F(r, z, \omega) = \omega S(r, \omega)\sin\theta\cos^{1/2}\theta\exp[i\omega(z\cos\theta/c - t)]$. For Equation (8), ultrafast hybridly vectorial-vortex light fields can be obviously degraded into hybridly vectorial light fields if we take $l = 0$ into account, so as to characterize ultrafast STO conversion process. While we are also able to flexibly adjust $l \neq 0$ for the purpose of clearly clarifying ultrafast OTLS conversion.

## 3. Ultrafast STO Conversion Within Hybridly Polarized Vectorial Light Fields

According to Equation (1) – (8) and Equation (S1) – (S2) (see supporting information (SI)), we will elaborate the ultrafast STO conversions, in which a hybridly polarized (and circularly polarized, as a comparison) LG femtosecond light pulse *in the absence of vortex phase* is incident into a high NA objective lens, thus leading to the ultrafast focused light fields. **Figure 2, 3** and Figure S1-S3 together with video 1 (see SI) show the corresponding ultrafast strongly focused light field distributions and associated phase textures, respectively. As showcased in Figure 2(a1)-2(d1), at the initial time (i.e., $t = 0$ fs), the total field (Figure 2(a1)), which can be regarded as the incoherent superimposition of its three-orthogonal component fields according to $|E_t|^2 = |E_x|^2 + |E_y|^2 + |E_z|^2$, exhibits doughnut-like-shaped pattern accompanied by four peripherally rotational symmetric hot spots. More importantly, its two transverse polarized component fields (Figure 2(b1) and 2(c1)), respectively, take on hollow suborbicular structures with two symmetrically bright points, wherein the distributions of two transverse component fields are perpendicular to each other. Such two intensity-null distributions in the center are capable to allow access to the existence of spatial phase singularity with spiral phase dislocation thus resulting in the azimuthal component of the Poynting vector that conduces to the generation of longitudinal OAM on $x$-$y$ planes, as shown in Figure 3(a1) and 3(b1), phenomenologically manifested by a phase step from 0 to $2\pi$ in an azimuthal cycle. As a result, STO conversion at the initial time occurs in two-orthogonal transverse component fields of hybridly polarized light field, instead of the longitudinal component field of the circularly polarized light fields, demonstrated in previous work [10]. This novel phenomenon might be conductive to rotating microparticles using the transverse fields rather than the longitudinal one. It is also interesting to note that a micro quadrotor aircraft appears within the longitudinal analogue (Figure 2(d1)) in which a hot spot sandwiched by four darker bright regions resides in the focus.

As time fleetingly elapsed ranging from 0 – 400 fs (the light fields at 400 fs not shown here are similar to that at 0 fs periodically), the overall field patterns (Figure 2(a1)-2(a8)) vary from the original smaller-sized non-zero hollow structure to the larger altogether null dark cores, following draw back to the started states (also see video 1). Both the dynamitic sizes and relative strengths with respect to such hollow regions at each time intervals are fully determined by that of its three-orthogonal component fields. While the time-variational light field distributions within a quasi-periodicity are dependent of the changeable time phase factor $\exp(i\omega t)$ according to Equation (8). It is paramount to pay attention to that two transverse component fields (Figure 2(b1)-2(b8) and Figure 2(c1)-2(c8)) carrying OAM of $\hbar$ per photon with mutually orthogonal patterns, respectively, change from a single ring into dual rings and then return to initial individual ones over ultrafast varying time, thus leading to the desired ultrafast STO conversion. As such, the corresponding ultrafast time-dependent phase textures (Figure 3(a1)-3(a8) and Figure 3(b1)-3(b8)) with spiral phase dislocations revolve incessantly, which are related with the time-assistant Gouy phase shift [41,42]. From closer inspection of these well-established time-varying phase structures, we can find that the phase sizes in the middle regions decrease with increasing ring numbers of light fields, and more remarkably detailed dynamic illustrations are vividly shown in video 1. For the sake of comparison, we also concern ultrafast STO conversions within pure circularly polarized light fields, as shown in Figure S1-S3 in SI. Distinguished from hybridly polarized light fields, the ultrafast conversions can always be reflected within longitudinal component fields for circularly polarized analogous. In addition, it is also observed from Figure 2(d1)-2(d8) that the center bright spot and four attached weak regions of the longitudinal counterparts, as time goes, start to be large and then they simultaneously shrink to the comparable sizes and at last also back to the original conditions. It is worth noting that there exist tiny energy exchanges with respect to the relative intensity between transverse and longitudinal component fields over ultrafast time, which are in principle ascribed to the redistributions of corresponding energy flux between longitudinal and transverse components.

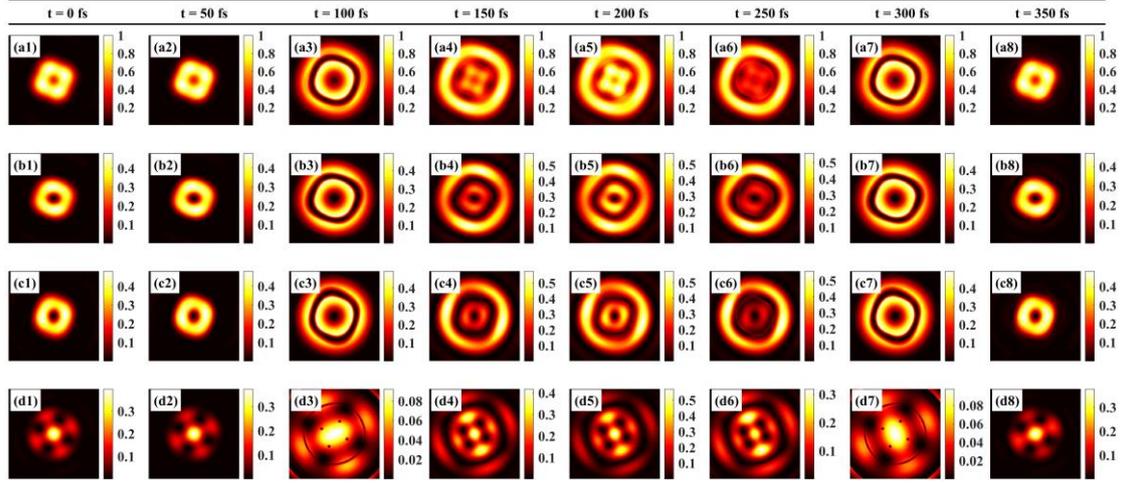

**Figure 2.** Ultrafast STO conversion within the two transverse component fields by tightly focusing hybridly polarized light pulse with argument setting $m = 1$, $n = 0$, $\varphi_0 = 0$, $l = 0$. The corresponding time-varying intensity profiles (on the $x$-$y$ planes) of total fields $|E_t|^2$ (a1-a8), transverse field components $|E_x|^2$ (b1-b8) and $|E_y|^2$ (c1-c8) as well as longitudinal field ones $|E_z|^2$ (d1-d8) at several optional time points of $t = 0$ fs, 50 fs, 100 fs, 150 fs, 200 fs, 250 fs, 300 fs, 350 fs. The simulated parameters throughout this paper are selected as follows. The laser pulse with a center wavelength of $\lambda = 633$ nm, central angular frequency $\omega_0 = 7.57 \times 10^{15}$ s$^{-1}$, beam size $\sigma_0 = 2$ cm, pulse width $T = 5$ fs. NA = 0.9. The color scale indicates the intensity in arbitrary units. The sizes for all of the images are $2\lambda \times 2\lambda$.

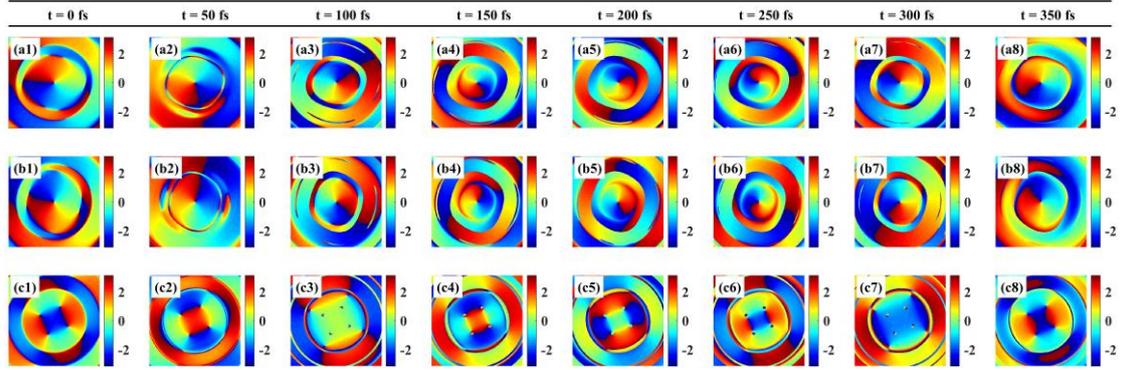

**Figure 3.** The temporal phase distributions on the transverse planes (i.e., $x$-$y$ planes) by tightly focusing hybridly polarized light pulse with argument setting $m = 1$, $n = 0$, $\varphi_0 = 0$, $l = 0$. The corresponding time-varying phase profiles of transverse component fields $|E_x|^2$ (a1-a8) and $|E_y|^2$ (b1-b8) and longitudinal ones $|E_z|^2$ (c1-c8) at several optional time points of $t = 0$ fs, 50 fs, 100 fs, 150 fs, 200 fs, 250 fs, 300 fs, 350 fs.

To further validate the ability of ultrafast STO conversion within transverse component fields of hybridly polarized light field, we adjust the incident TC from $l = 0$ to $l = 1$ while keeping the other arguments constant for the purpose of the expectation of the resultant OAM of $2\hbar$ per photon within two transverse component fields, respectively. **Figure 4, 5** together with video 2 jointly illustrate the ultrafast light field distributions and the associated phase textures of hybrid polarized (including linear, circular, elliptical polarizations) vortex light pulse with $m = 1$, $n = 0$, $\varphi_0 = 0$, $l = 1$. It can be seen from Figure 4(a1) that at the static time interval (i.e., $t = 0$ fs), the overall field presents elliptical bright spots distribution. Unfortunately, the hollow light field structure exists only within *x*- component field (Figure 4(b1)), while the transverse *y*-component field (Figure 4(c1)) shows the unexpected center hot spot pattern, which is due mainly to the influence of the helicity of incident spiral phase front (i.e., the sign of the incident TC) upon polarized light field components, rooted in that the polarization symmetry of vector light field is broken (the detailed illustrations are shown in SI in section 3). Remarkably, we are capable to clearly verify again the STO conversation existed in transverse component field (*x*-component field) from the phase textures, as shown in Figure 5(a1), in which there exist double screw dislocations manifested mathematically by $\exp(2i\varphi)$ where one phase step of $0$-$2\pi$ derives from incident TC of $l = 1$ and the other roots from desired STO conversion. By the way, the dipole-like-shaped hot points tilted versus *x* direction appears in the longitudinal component field, as shown in Figure 4(d1).

As time elapses, the total field patterns (Figure 4(a1)-4(a8)) vary from the initial elliptical hot spot into sandwich structure and then draw back to the beginning case. It can be observed that the hollow region of *x*- component fields (Figure 4(b1)-4(b8)) enlarge with varying time and the corresponding phase textures (Figure 5(a1)-5(a8)) with dual screw dislocations also revolve constantly, which means that there indeed exist the ultrafast STO conversions within hybridly polarized light fields. Beyond that, the alternation between bright spots and dark cores arises within *y*- component fields as the time fleetingly changes, as shown in Figure 4(c1)-4(c8) and detailly illustrated in video 2. In principle, such light field structures can be attributable to the inevitable dependence on the temporal phase factor and constructive and destructive interferences between transverse and longitudinal component fields, thus which might conduce to the trapping and manipulation of the charged particles with two types of refractive indexes with respect to ambient [43-45]. As such, the corresponding ultrafast time-varying phase distributions are detailed illustrated in Figure 5(b1)-5(b8) and video 2, respectively. Incidentally, the longitudinal ones (Figure 4(d1)-4(d8)) vary from tilted single semi-lune to the double counterparts over ultrafast time.

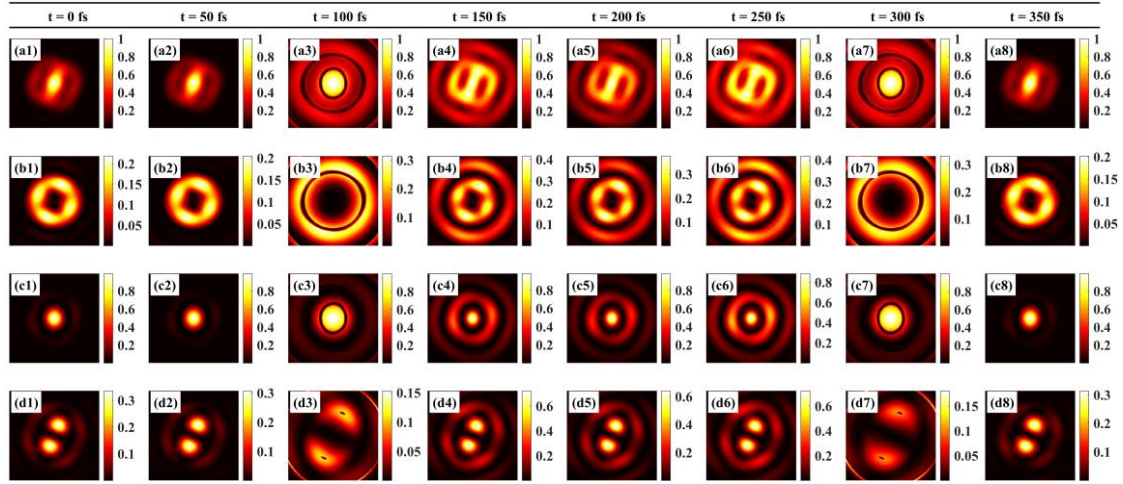

**Figure 4.** Ultrafast STO conversion within *x*- component fields by tightly focusing hybridly polarized light pulse with argument setting $m = 1$, $n = 0$, $\varphi_0 = 0$, $l = 1$. The corresponding time-varying intensity profiles of total fields (a1-a8), transverse field components (b1-b8) and (c1-c8) as well as longitudinal field components (d1-d8) at several optional time points of $t = 0$ fs, 50 fs, 100 fs, 150 fs, 200 fs, 250 fs, 300 fs, 350 fs.

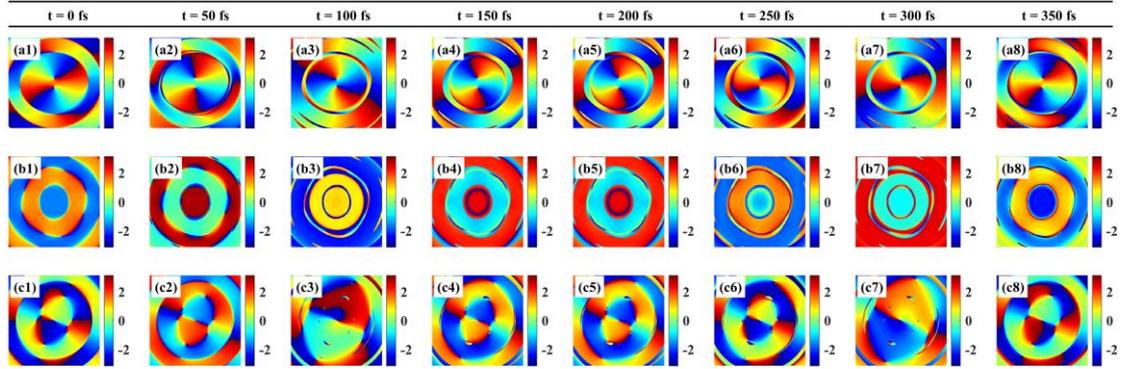

**Figure 5.** The temporal phase distributions on the transverse planes (i.e., *x-y* planes) by tightly focusing hybridly polarized light pulse with argument setting $m = 1$, $n = 0$, $\varphi_0 = 0$, $l = 1$. The corresponding time-varying phase profiles of transverse field components (a1-a8) and (b1-b8) and longitudinal field components (c1-c8) at several optional time points of $t = 0$ fs, 50 fs, 100 fs, 150 fs, 200 fs, 250 fs, 300 fs, 350 fs.

## 4. Ultrafast OTLS Conversion Within Hybridly Polarized Vectorial-Vortex Light Fields

In the next step, we will turn to proceed to the ultrafast OTLS conversion by exploiting tight focusing hybridly polarized LG femtosecond light pulse *in the presence of vortex phase* with $m = 0$, $n = 0$, $\varphi_0 = 0$, $l = 2$ (i.e., including only linear polarization) and azimuthally polarized counterparts as a comparison (see SI in section 2). **Figure 6** and Figure S4 together with video 3, respectively, graphically depict the tightly focused light field distributions superimposed with the polarization maps and the resultant ultrafast three-orthogonal SAM density components. The time-dependent SAM density can be described by the following formulation [46],

$$\mathbf{S} = \frac{\mathrm{Im}\left[\varepsilon\left(\mathbf{E}^* \times \mathbf{E}\right) + \mu\left(\mathbf{H}^* \times \mathbf{H}\right)\right]}{4\omega}, \tag{9}$$

where $\varepsilon$ and $\mu$ are, respectively, permittivity and permeability in the propagating medium. Considering the femtosecond light pulse propagating through the free space which is a non-magnetic medium, we here focus only on the electric fields, i.e., neglecting the second term shown in Equation (9). Hence, the resulting three SAM density components can be derived as,

$$S_x = \frac{\varepsilon \operatorname{Im}\left[E_y^* E_z - E_z^* E_y\right]}{4\omega} = \frac{\varepsilon |E_y||E_z|\sin(\varphi_z - \varphi_y)}{2\omega},$$

$$S_y = \frac{\varepsilon \operatorname{Im}\left[E_z^* E_x - E_x^* E_z\right]}{4\omega} = \frac{\varepsilon |E_z||E_x|\sin(\varphi_z - \varphi_x)}{2\omega}, \quad (10)$$

$$S_z = \frac{\varepsilon \operatorname{Im}\left[E_x^* E_y - E_y^* E_x\right]}{4\omega} = \frac{\varepsilon |E_x||E_y|\sin(\varphi_y - \varphi_x)}{2\omega},$$

in which $\varphi_j$ stand for the phase manifested by $\operatorname{Arg}(E_j)$ within tightly focused light fields.

As shown in Figure 6(a1), at $t = 0$ fs, the total field presents hollow obliquely ellipse patten along the long axis of 45 degrees versus the positive $x$ direction (-45 degrees by adjusting $\varphi_0$ from 0 to $\pi/2$), while a pair of symmetrical bright spots appear on the short axis of the ellipse pattern. Most importantly, the locally circular polarization at the central focus is able to be induced, which implies that there exists the OTLS conversion within hybridly polarized vortex light field. Besides, the resultant SAM density components at initial time are shown in Figure 6(b1)-6(d1), in which each component is normalized into its maximum, respectively. As shown in Fig. 6, the distributions of SAM components have pseudo two-fold rotational symmetry. It is worth noting that the total SAM remains the conservation, and their quantities are always equal to zero, which means that the SAM-OAM conversion occurs locally. As time elapses, it is observed from Figure 6(a1)-6(a8) that the hollow regions of the total fields gradually enlarge at $t = 100$ fs, and then go back to its original state. In the meantime, the superposed polarization maps in the periphery also embody the dependence on ultrafast fleeting time (see video 3), which might have profound implications for the time-associated polarization. However, the central locally circular polarization always remains constant, thus leading to the steady ultrafast OTLS conversions. In order to concern the orbit-induced local SAM over time, we proceed to plot the patterns of three orthogonal SAM density components (Figure 6(b)-6(d)) at the corresponding time intervals. As depicted, the resultant SAM components experience alternative changes between bright and dark fields, i.e., appearance and annihilation. This alternation of SAM density components can be attributed to the constructive and destructive interference among distinct temporal light field components resulting from time-varying phase factors. Therefore, altering intrinsic ultrafast time embedded into the light pulses provides a crucial knob to manipulate the local SAM density within the focused light fields.

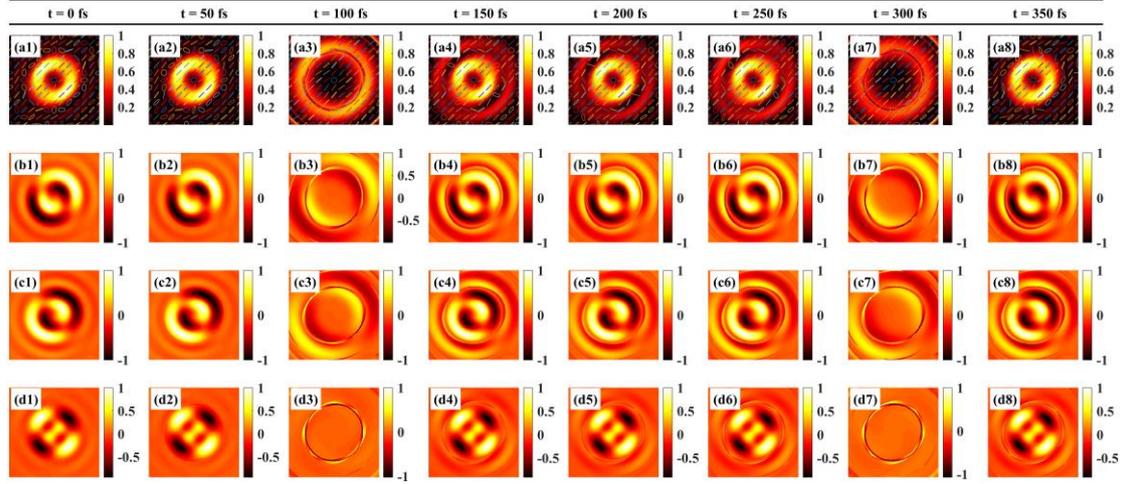

**Figure 6.** (a1-a8) Ultrafast OTLS conversion within total light fields superimposed with polarization maps of tightly focused hybidly polarized light beams with $m = 0$, $n = 0$, $\varphi_0 = 0$, $l = 2$. The resulting time-varying SAM profiles of $x$- components (b1-b8), $y$- components (c1-c8) and $z$- components (d1-d8) at several optional time points of $t = 0$ fs, 50 fs, 100 fs, 150 fs, 200 fs, 250 fs, 300 fs, 350 fs.

For the purpose of the further analysis, we explore the influence of various incident TC ranging from $l = 0$ to 3 under distinct time intervals from $t = 0$ to 400 fs on the locally induced three-orthogonal SAM density components, respectively. The outcomes are shown in **Figure 7**. Firstly, we can see that at $t = 0$ fs, two transverse SAM density (Figure 7(a1) and 7(b1)) of $S_x$ and $S_y$ always exist regardless of the existence of TC, whereas only when there exists TC, the OTLS occurs in $S_z$ (Figure 7(c1)). This is due mainly to that the Stokes parameter $S_3$, proportional to $S_z$, is null, thus leading to there is no induced local polarization at $l = 0$. From the transverse SAM density, we can first find that the reverse directions between $S_x$ and $S_y$ are always on the opposite. As an example of the $S_x$, the value of TC also affects the reverse direction from negative ($l = 0$) to positive ($l > 0$). As such, the SAM gradually goes away from the center with the increase of OAM, and the value of TC also has the influence on the magnitude of the transverse SAM density, namely, the maximum is slightly upward with the increasing TC. In contrast, TC appreciably affects the reverse direction increasing TC from $l = 2$ to 3 in the longitudinal SAM density. Therefore, to cherry-pick the SAM density components and the values of OAM provides the powerful knobs for the purpose of the manipulation of the local SAM density within the tightly focused fields.

And then, as time elapses, the distributions of the transverse SAM density (Figure 7(a1)-7(a5) and 7(b1)-7(b5)) vary periodically from 0 fs to 400 fs, regardless of the existence of incident OAM. This is mainly owing to periodically time-varying phase factor $\exp(i\omega t)$, thus leading to time-dependent SAM density. Similarly, the varying trends of transverse SAM density components at $t = 0$ fs under each TCs are consistent with that at $t = 400$ fs. It also can be found from the transverse SAM density that, there exist noticeable side lobes upon the off-axis positions under distinct time intervals. As a comparison, the longitudinal ones (Figure 7(c1)-7(c5)) fail to experience precise periodical distributions (especially with $l = 0$) with fleetingly elapsed time. This is attribute to the strong dependence on the values of OAM and the ultrafast temporal effect. More noticeably, the null $S_z$ with TC of $l = 0$ at $t = 0$ fs, in which the incident hybridly polarized light pulses only include linear polarization which means that the term $Im(\mathbf{E}\times\mathbf{E}^*)= 0$, can spontaneously evolve into nonzero distributions as time goes by. Such novel phenomenon stems from that the time $t$ affects the intrinsic phase distributions of light pulse, thus giving rise to the time-varying phase difference between each SAM density components following Equation (10). This effect might be conducive to further exploring the ambiguous time polarization with highly desired functionalities to perceive more new effects [47,48]. As a consequence, this novel ultrafast time degree of freedom proposed here indeed can be view an alternative way to tailor and harness desirous SAM density fields.

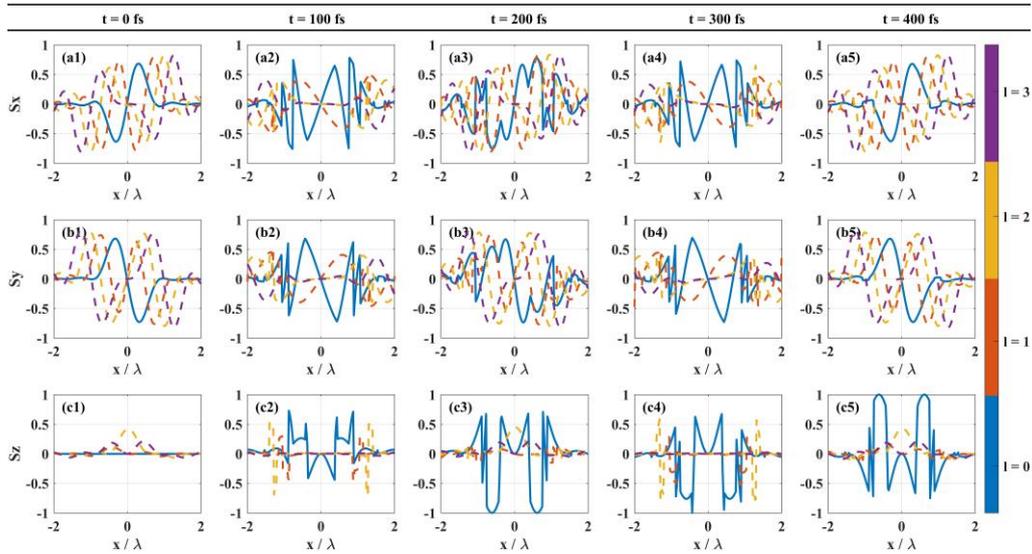

**Figure 7.** The dependence of the various TC of OAM on three-orthogonal SAM density components along the *x* axis, including *x*- components (a1-b5), *y*- components (b1-b5) and *z*- components (c1-c5) under several optional time points of $t = 0$ fs, $t = 100$ fs, $t = 200$ fs, $t = 300$ fs, $t = 400$ fs. The color bar located in the right side represents the line color of the distinct TC from 0 to 3. The solid blue lines corresponding to TC of $l = 0$. The dotted lines with orange, yellow and purple colors corresponding to TC of $l =$1, 2, 3, respectively.

## 5. Conclusions

In summary, we have proposed and demonstrated theoretically the ultrafast interconversions between SAM and OAM by tightly focused hybridly polarized light pulses, based on the time-associated vectorial diffraction theory and the fast Fourier transform. Firstly, at the initial time, we find that incident hybridly polarized vectorial light pulses (including linear, circular, elliptical polarizations) empower transverse component fields with order-adjustable OAM, therefore implying the existence of controllable STO conversion. Reversely, hybridly polarized vectorial-vortex light pulses (including only linear polarization) allow access to the locally induced circular polarization at the focus, thus indicating the emergence of tunable OTLS conversion. As a result, spin-orbit reciprocal conversion occurs at the static time. Secondly, as time elapses, it is unveiled that there exist ultrafast STO conversion processes, wherein the converted light fields with resultant OAM remain constantly rotating. Principally, such rotational OAM-carrying fields can be attributed to time-related Gouy phase shift. In a reversable manner, ultrafast OTLS conversion comes up, in which the locally induced circular polarizations at the central focus of the total fields are independent of the ultrafast varying time, while ultrafast temporal effect endows three-orthogonal SAM density components with the alternation between the appearance and annihilation. This alternation of SAM density components owes to the constructive and destructive interference among distinct temporal light field components resulting from time-varying phase factors. As a consequence, ultrafast spin-orbit reciprocal conversion occurs within hybridly polarized light fields. Such proposed methodology and demonstrated findings not only provide a novel degree of freedom into the spin-orbit interaction of light, but also contribute to the possible applications in optical micro-manipulation and micro-fabrication.

**Supporting Information**

Supporting Information is available from the Wiley Online Library or from the author.

**Acknowledgements**

This work was supported by the National Natural Science Foundation of China (Nos. 62075049, 11974258, 11604236, 61575139), Key Research and Development (R&D) Projects of Shanxi Province (201903D121127), Scientific and Technological Innovation Programs of Higher Education Institutions in Shanxi (2019L0151), and the Natural Sciences Foundation in Shanxi Province (201901D111117).

**Conflict of Interest**

The author declares no conflict of interest.

**Keywords:** ultrafast optical field, spin-orbit coupling, vectorial diffraction theory, fast Fourier transform, vectorial light beam

Received: ((will be filled in by the editorial staff))
Revised: ((will be filled in by the editorial staff))
Published online: ((will be filled in by the editorial staff))